\documentclass[12pt,preprint]{aastex}

\shorttitle{Rubidium in the Interstellar Medium}
\shortauthors{Walker et al.}

\begin{document}

\title{Rubidium in the Interstellar Medium}

\author{Kyle M. Walker\altaffilmark{1,2}, S. R. Federman\altaffilmark{1,2},
David C. Knauth\altaffilmark{3,4}, and David L. Lambert\altaffilmark{5}}

\altaffiltext{1}{Department of Physics and Astronomy, University of Toledo,
Toledo, OH 43606; kwalker@physics.utoledo.edu; steven.federman@utoledo.edu.}
\altaffiltext{2}{Guest Observer, W.J. McDonald Observatory, University of Texas 
at Austin, Austin, TX 78712.}
\altaffiltext{3}{Franklin High School, Reisterstown, MD 21136; 
knauth$\_$dc2@hotmail.com.}
\altaffiltext{4}{Visiting Scientist, Department of Physics and Astronomy, 
Johns Hopkins University, Baltimore, MD 21218}
\altaffiltext{5}{W.J. McDonald Observatory, University of Texas at Austin, 
Austin, TX 78712; dll@astro.as.utexas.edu.}

\begin{abstract}
We present observations of interstellar rubidium 
toward $o$ Per, $\zeta$~Per, AE Aur, HD 147889, 
$\chi$ Oph, $\zeta$ Oph, and 20 Aql. 
Theory suggests that stable $^{85}$Rb and long-lived $^{87}$Rb are produced 
predominantly by high-mass stars, through a combination of the weak s-
and r-processes.
The $^{85}$Rb/$^{87}$Rb ratio was determined from measurements of
the Rb~{\small I} line at 7800 \AA\ and was compared 
to the solar system meteoritic ratio of 2.59.
Within 1-$\sigma$ uncertainties all directions except HD 147889 have Rb 
isotope ratios consistent with the solar system value.
The ratio toward HD 147889 is much lower than the meteoritic 
value and similar to that toward $\rho$ Oph A \citep{fed04}; 
both lines of sight probe the Rho Ophiuchus Molecular Cloud.
The earlier result was attributed to a deficit of r-processed $^{85}$Rb.
Our larger sample suggests instead that $^{87}$Rb is enhanced in these
two lines of sight. 
When the total elemental abundance of Rb is compared to the K elemental
abundance, the interstellar Rb/K ratio is significantly lower than the 
meteoritic ratio for all the sight lines in this study.
Available interstellar samples for other s- and r- process elements 
are used to help interpret these results.
\end{abstract}

\keywords{ISM: abundances --- ISM: atoms --- stars: individual ($o$ Persei, 
$\zeta$ Persei, AE Aurigae, HD 147889, $\chi$ Ophiuchi, $\zeta$ Ophiuchi, 
20 Aquilae)}

\section{Introduction}

Rubidium isotopes $^{85}$Rb and $^{87}$Rb are products of the neutron 
capture slow (s-) and rapid (r-) processes. ($^{87}$Rb is unstable 
with the long halflife of 4.75 $\times$ 10$^{10}$ years.) The s-process 
contributions come from two types of stars: the so-called main s-process 
operates in low mass asymptotic giant branch (AGB) stars with their products 
ejected into the 
interstellar medium (ISM) by the stellar wind; and the weak s-process 
occurs in the He- and C-burning shells of massive stars and distributed 
into the ISM primarily, it is thought, by a star's terminal supernova 
explosion. The r-process most likely occurs at the time of the 
supernova explosion. Of primary relevance to studies of the Rb elemental 
and isotopic abundances beyond the solar system is the division of 
responsbility for Rb synthesis between low-mass long-lived and high-mass 
short-lived stars or, equivalently, the relative contributions from the 
main s-process and the combination of the weak s-process and the r-process. 

Dissection of the solar system abundances for Rb and 
adjacent elements enables fractional contributions of the 
weak and main s-processes to be estimated. Then, the r-process 
contribution is obtained as the differences between the solar system 
abundance and the sum of the weak and main s-process contributions. 
The most recent exercise of this kind is that reported by \citet{heil} 
who estimated the fractional contributions of the main s-process 
to the solar system abundances of $^{85}$Rb and $^{87}$Rb to be 
0.17 and 0.24, respectively. In other words, the relative 
contributions of low mass and high mass stars 
is 0.17/0.83 = 0.20 for $^{85}$Rb 
and 0.24/0.76 = 0.32 for $^{87}$Rb. [\citet{heil} give the weak s-process 
fractional contributions as 0.24 for $^{85}$Rb and 0.46 for $^{87}$Rb and, then 
by subtraction of the two s-process contributions, the r-process 
contributions are 0.59 for $^{85}$Rb and 0.30 for $^{87}$Rb.] 
Earlier dissections of the solar system abundances have given 
somewhat different results for the relative contributions 
of low and high mass stars, e.g., \citet{arland} gave 
0.16/0.84 = 0.19 for $^{85}$Rb and 0.35/0.65 = 0.54 for $^{87}$Rb indicating 
a more pronounced role of the massive stars in the synthesis of $^{87}$Rb 
(relative to $^{85}$Rb) than suggested by \citet{heil}.
Most probably, higher weight should be given to Heil et al.'s result because 
it was based in part on new and accurate measurements of neutron 
capture cross sections, especially for the two Rb isotopes. 
Our study of interstellar Rb was undertaken to set observational 
constraints on the proposed mechanisms of Rb nucleosynthesis.

Rubidium has been analyzed in both unevolved and evolved stars.
\citet{lluck} sought the Rb isotope ratio in Arcturus.
\citet{gratton} studied the abundances of neutron-rich elements in 
metal-poor stars, while \citet{tomkin} considered rubidium in 
metal-deficient disk and halo stars.
Others have examined rubidium abundances in AGB stars 
\citep{lambert,abia2,abia1}.
However, rubidium is difficult to measure in stellar atmospheres.
The Rb~{\small I} lines at 7800 and 7947 \AA\ are blended with nearby lines
and stellar broadening prevents the hyperfine and isotopic components 
from being resolved.
Intrinsically narrow interstellar lines are not hindered by these 
complications.

The detection of rubidium in the ISM was first reported
by \citet{jura} from absorption toward the bright star $\zeta$ Oph.
They used the Rb~{\small I} 
5{\it s} $^{2}${\it S}$_{1/2}$--5{\it p}~$^{2}${\it P}$_{3/2}$ 
line at 7800.29 \AA.
This is the strongest resonance line of neutral rubidium 
and it is used for our study as well, even though most rubidium in 
diffuse interstellar clouds is ionized because it has a low ionization 
potential of 4.177 eV. 
Rubidium's overall low abundance requires the use of its strongest line.

The early detection by \citet{jura} could not be confirmed by 
\citet{fed85}, who also could not detect Rb~{\small I} 
toward $o$ Per nor $\zeta$ Per. 
\citet{fed85} calculated a typical 3-$\sigma$ upper 
limit in column density of $\sim$ 5 $\times$ 10$^{9}$ cm$^{-2}$ for 
these lines of sight.
Interstellar absorption toward these three stars is revisited in 
this study.
More recently, \citet{gredel} provided the first firm detections of 
interstellar rubidium toward Cyg OB2 No. 12 and No. 5. 

The solar system isotope ratio is found from carbonaceous chondrite 
meteorites, with a ratio of $^{85}$Rb/$^{87}$Rb = 2.59 \citep{lodders}.
\citet{fed04} found an interstellar isotope ratio toward $\rho$ Oph A 
of 1.21, which differs significantly from the meteoritic value. 
This was the first relatively precise determination of the rubidium isotope 
ratio for extrasolar gas.  In a recent paper, Kawanomoto et al. 
(2009) studied the line of sight toward HD~169454 and derived an 
isotope ratio that is consistent with the solar system value.  
The present study expands upon the results of \citet{fed04} and 
Kawanomoto et al. (2009) on the stable $^{85}$Rb and long-lived 
$^{87}$Rb isotopes.

We sought Rb absorption along seven lines of sight toward $o$ Per, 
$\zeta$ Per, AE Aur, HD 147889, $\chi$ Oph, $\zeta$ Oph, and 20 Aql.
These lines of sight were selected because they sample different directions in
the solar neighborhood, and because other relevant data noted below are 
available for them.
Furthermore, the cloud structure is simple with only one or two 
apparent components per line of sight. 
We explore $^{85}$Rb/$^{87}$Rb ratios and compare the results to those
for $\rho$~Oph~A. 
We also compare the elemental abundance ratio 
of Rb/K, as well as the $^{85}$Rb/K and $^{87}$Rb/K ratios, to examine 
whether the $^{85}$Rb or $^{87}$Rb abundances are lower or higher 
relative to the abundances seen in meteorites.
Section 2 summarizes the high-resolution observations and data 
reduction techniques on the Rb~{\small I} $\lambda$7800 line. 
Section 3 presents the analysis of the data and Section 4 the results, 
including confidence limits on the column densities.  
Section 5 uses the results to discuss the implications for interstellar 
rubidium and its nucleosynthetic production site(s).  
Finally, Section 6 gives a summary and conclusion. 

\section{Observations and Data Reduction}

We observed the stars with the Robert G. Tull Spectrograph -- formerly 
2dcoud\'{e} spectrograph -- 
\citep{tull} on the Harlan J. Smith 2.7 m telescope at McDonald
Observatory from 2005 to 2007.  
We used the high-resolution mode with echelle grating E1 to acquire 
spectra on a 2048 $\times$ 2048 Tektronix CCD. 
Depending on the central wavelength, there were either 12 or 13 orders  
imaged onto the CCD and each order was around 30 \AA\ wide. 
The grating was centered near 7298 \AA\ to capture the Rb~{\small I} 
$\lambda$7800 feature in either the eighth or ninth order.  
A 145 $\mu$m slit gave a resolving power of 175,000 (2.9 pixels per 
resolution element), as determined from the width of lines in the 
Th-Ar comparison spectra.  The resolving power varied from 172,000 to 
184,000 from night to night and from epoch to epoch.  
For calibration, we obtained bias and flat-field exposures at the 
beginning of each night.
Furthermore, we acquired a set of 30-minute long dark current 
exposures on the first night of the observing runs.
The dark images were used to 
determine the rate that thermal electrons accumulated on the CCD. 
Every 2-3 hours throughout the night we obtained a Th-Ar 
comparison spectrum for wavelength calibration.
Also each night we observed an unreddened star such as $\alpha$ 
Leo or $\gamma$ Cas centered on the slit to see if there were any
CCD blemishes not removed by the flat-fielding process.  
There was a seven-pixel wide CCD detector glitch in the 2005 December run 
that affected the spectra of $\zeta$ Per.  
We were unable to remove the glitch and subsequently
those spectra were not used.  
In all other runs the glitch did not fall on the Rb~{\small I} feature. 
This was due to the fact that we shifted the center position of the
detector slightly each night to minimize the chance of the feature falling
on corrupted pixels.
Individual exposures were limited to 30$^m$ to minimize the 
presence of cosmic rays.  
Table 1 lists the stellar and observational parameters: HD number, 
name, spectral type, {\it V} magnitude, and 
distance in pc (obtained from the SIMBAD
database, operated at CDS, Strasbourg, France), dates observed, 
exposure time, and unreddened stars observed.


The data were reduced using standard routines within the {\small IRAF} 
environment.
Master bias and dark images were subtracted from all the raw object, 
comparison, and flat images.
These images were processed further to apply the overscan strip correction. 
Cosmic rays were eliminated in the object and comparison images and 
these images, along with the master flat, were fit with a low-order 
polynomial that subtracted the scattered light from each order. 
The master flat was normalized to unity and divided into each object 
and comparison image and the light from each aperture was summed to
form one-dimensional spectra.
At this point, the spectra of the program and unreddened 
stars were compared to check for CCD glitches or artifacts.
Typically three or more Th-Ar lines in the comparison spectra were identified 
per order for the wavelength solution.  The rms deviation averaged 10$^{-4}$ 
\AA\ by fitting the x and y directions to second order polynomials. Over 
the entire spectrum, there were few moderately strong lines.
Two comparison spectra were assigned to each spectrum and 
it was corrected for dispersion.
Each spectrum was then Doppler-corrected and all were summed
to obtain a spectrum for each night.  
Each night's summed spectra were compared to make sure there were no 
discrepancies in the width or depth of the Rb~{\small I} feature.  
None were present.
Finally, the individual exposures for each line of sight were summed
to produce seven spectra.

Table 2 gives the signal-to-noise ratio per pixel and the 
equivalent width ({\it W}$_\lambda$) of the line divided by its 
uncertainty ($\sigma$). 
The equivalent width was measured using the deblending function in 
{\small SPLOT} of {\small IRAF}, where Voigt profiles were fit to the features.
The uncertainty was calculated by dividing the full width at half
maximum (FWHM) of the Rb~{\small I} feature by the average 
signal-to-noise (S/N) in the stellar continuum. 
The goal was to achieve values for {\it W}$_\lambda$/$\sigma$ of 
about 10 so that the Rb~{\small I} profiles could be fit with confidence.


\section{Analysis}

The Rb~{\small I} spectra were synthesized using a C++ program 
(J. Zsarg\'{o}, modified by Knauth, hereafter called RbFits) to fit
the profile, taking the instrumental width into account.
RbFits took as input {\it V}$_{LSR}$, the {\it b}-value, and 
the column density, {\it N}.
In order to determine the starting input velocities and {\it b}-values 
for the RbFits program, K~{\small I} parameters were used, as K~{\small I} 
is expected to trace Rb~{\small I} in diffuse interstellar gas.
Whenever possible, results from the weak K~{\small I} lines at 4044 and
4047 \AA, whose values of {\it W}$_\lambda$ are similar to those of Rb~{\small I},
were adopted.
Occasionally the strong line at $\lambda$7699 was used to find the component
structure, but in all cases the weak lines provided K~{\small I} column
densities.
It is also worth noting that the K~{\small I} lines have unresolved hyperfine 
components and are not affected by the presence of isotopes; thus their
profiles are relatively simple.
In fact, all the alkalis are likely to coexist in diffuse interstellar 
clouds \citep{welty,knauth03,fed04}; if the K~{\small I} 
did not trace Rb~{\small I} particularly well, $^{7}$Li~{\small I} and 
$^{6}$Li~{\small I} were used as a guide.
The rubidium absorption toward $\chi$ Oph and 20 Aql incorporated
starting values from the Li~{\small I} data;
all other lines of sight began their fits with K~{\small I} velocities 
and {\it b}-values.
The velocities reported below are based on fits to the Rb~{\small I} profile.
These final Rb~{\small I} velocities agree well with those determined for 
Li~{\small I} and K~{\small I} \citep[e.g.,][]{knauth03}.
The RbFits program fit Voigt profiles to the features taking the 
hyperfine structure of both $^{85}$Rb~{\small I} and $^{87}$Rb~{\small I} 
into account. 
The wavelengths and {\it f}-values from \citet{morton} are given in Table 3.


When fitting the profile, RbFits calculated the lowest 
$\chi$$^{2}$ by modifying the input values. 
Also, each input parameter of each cloud was either allowed to be held
fixed or to vary.
The goal was to obtain a synthesized profile where the velocity and 
{\it b}-values for both isotopes were equal, and the column densities of 
each cloud formed an isotope ratio at least similar to the solar system 
value, 2.59.
The inferred isotope ratio did not depend on the initial value of 2.0 or 2.5.
At the same time, a check was performed on the reduced $\chi$$^{2}$, which
ideally should be $\sim$1.0.
For lines of sight with one cloud, there were 43 degrees of freedom; there 
were 51 degrees of freedom for lines of sight with two clouds (one degree 
per data point of the feature minus the number of parameters being fit).

The simplest lines of sight were those that had only one cloud component, 
$\zeta$ Per, HD147889, and $\chi$ Oph.
The K~{\small I} template for $\zeta$ Per was taken from \citet{knauth00},
and that for $\chi$ Oph was from \citet{knauth03}.
An HD 147889 K~{\small I} spectrum for the line at 7699 \AA\ was obtained 
from our unpublished results and was fitted using a {\small FORTRAN} 
InterStellar MODelling program, {\small ISMOD} (Sheffer, unpublished), 
because there was no cloud structure available in the literature. 
Using velocities, {\it b}-values, and equivalent widths estimated from 
{\small IRAF} routines, {\small ISMOD} fit a profile to the HD 147889 
data that included unresolved hyperfine structure.
The results from the main component were used as the starting point 
for RbFits.

\citet{fed04} performed a synthesis with RbFits on $\rho$ Oph A. 
The first step in using the RbFits program was to reproduce the results 
from that study.
First, all six parameters were allowed to vary.
From this output, both velocities were set to 1.96 km s$^{-1}$ and both 
{\it b}-values were set to 1.43 km s$^{-1}$ from the solution of $^{87}$Rb.
RbFits was run again with these four values held fixed, only allowing the 
column density to vary.
The resulting solution had a reduced $\chi$${^2}$ of 1.57, similar to the 
1.43 of \citet{fed04}.
The $^{87}$Rb/$^{85}$Rb isotope ratio for $\rho$ Oph A was computed to be 
1.22 $\pm$ 0.30, consistent with the 1.21 $\pm$ 0.30 obtained by 
\citet{fed04}.
Furthermore, a plot of the spectrum of $\rho$ Oph A was created with the 
data and fit.
A visual inspection was performed to make sure that the 
residuals (data minus fit) were the same inside and outside the profile. 

The above technique was used to obtain syntheses for the seven new lines 
of sight. 
The basic routine was to start by varying everything and then fixing 
more and more parameters until the solution had a reduced $\chi$${^2}$ 
around 1.0, the residuals were consistent along the profile, and the 
$^{87}$Rb/$^{85}$Rb isotope ratio was reasonable.

\subsection{One Component Fits}

The starting input values for $\chi$ Oph were taken from the K~{\small I} and
$^{7}$Li~{\small I} values of \citet{knauth03}.
The input velocity and {\it b}-value were 0.20 and 1.00 km s$^{-1}$, 
respectively.
Because the lines are so weak, the input column density was derived 
from the ratio of equivalent width to column 
density for $\rho$ Oph A using the simple equation

\begin{equation}
\frac{W{_\lambda}({\rm \rho\ Oph\ A})}{{\it N}({\rm \rho\ Oph\ A})} = 
\frac{W{_\lambda}({\rm \chi\ Oph})}{{\it N}({\rm \chi\ Oph})},
\end{equation}

\noindent where {\it W}$_\lambda$(X) is the equivalent width of the 
Rb~{\small I} line and {\it N} is the total column density of $^{87}$Rb 
and $^{85}$Rb combined.
Once the total column density was determined, it was divided so that the 
isotope ratio would be similar to that of the solar system. 
Thus, 67\% of {\it N}({$\chi$ Oph) was $^{85}$Rb and 33\% of it was 
$^{87}$Rb, using a ratio of 2:1 that is midway between the results for
$\rho$ Oph A \citep{fed04} and the solar system \citep{lodders}. 
These were the column densities used for the starting point in RbFits. 
$\chi$ Oph has the highest reduced $\chi$${^2}$, but the synthesized 
profile fits the data well.
This ``ratio'' technique was also used for HD 147889.

The profile for $\zeta$ Per was synthesized in three steps, following the 
above procedure. 
The starting values were taken from \citet{knauth00}.
During the second step, however, instead of using the velocity output 
from $^{87}$Rb~{\small I}, the velocity of $^{85}$Rb~{\small I} was more 
similar to that of K~{\small I} and $^{7}$Li~{\small I} from \citet{knauth00}.
After using the velocity and {\it b}-value from the solution of 
$^{85}$Rb~{\small I}, the rest of the synthesis was performed as described 
above.
Figure 1 displays the fitted spectra toward $\zeta$ Per, $\chi$ Oph, 
and HD 147889.


\subsection{Two Component Fits}

The profiles of the four lines of sight with two components were more 
difficult to synthesize.  The starting values did not affect the 
outcome of the one-component fits, whereas 
the simple ``ratio'' technique of dividing equivalent widths by column 
densities was no longer valid because the profiles were more complex,
each having eight components (two hyperfine components for each 
isotope for each cloud).
The program needed better starting values from other data 
(velocity separations, etc.), especially for the column densities.
There are two ways to solve this problem.

For the first way, a list of abundances of Rb~{\small I} and K~{\small I} 
was created from lines of sight with one component: $\chi$ Oph, $\zeta$ Per, HD 
147889, $\rho$ Oph A, and the two lines of sight from \citet{gredel},
Cyg OB No. 5 and No. 12.
The K~{\small I} column densities were obtained from \citet{knauth00} 
($\zeta$ Per), \citet{knauth03} ($\chi$ Oph), 
\citet{fed04} ($\rho$ Oph A, Cyg OB No. 5 and No. 12), and derived from 
unpublished observations (HD~147889).
An average of the {\it N}(Rb~{\small I})/{\it N}(K~{\small I}) ratios for 
each line of sight yielded 9.5 $\times$ 10$^{-4}$.
At this preliminary stage of the analysis for AE Aur, the component structure 
and {\it N}(K~{\small I}) come from 
the synthesis of $\lambda$7699 from unpublished data.
For all other directions, {\it N}(K~{\small I}) is obtained 
from the weak line at $\lambda$4044.
Using the above average and knowing K~{\small I} for the four lines of sight in
the current study, one could work backwards to find {\it N}(Rb~{\small I}).
Then using {\it N}(Rb~{\small I}) as the total column density for a 
given line of sight and the cloud fractions from \citet{knauth03} for $o$ Per,
$\zeta$ Oph, and 20 Aql as well as our results for AE Aur, the column densities 
for each cloud were obtained.

For example, $o$ Per has a total K~{\small I} column density of 1.04 
$\times$ 10$^{12}$ cm$^{-2}$. Then, according to \citet{fed04} and assuming
equal depletion of Rb and K although the elements have somewhat different 
condensation temperatures \citep{lodders},

\begin{equation}
{\Bigg[}\frac{A{_g}({\rm Rb})}{A{_g}({\rm K})}{\Bigg]} = 
{\Bigg[}\frac{{\it N}({\rm Rb\ {\small I}})}{{\it N}({\rm K\ {\small I}})}{\Bigg]} 
{\Bigg[}\frac{G({\rm Rb\ {\small I}})}{G({\rm K\ {\small I}})}{\Bigg]} 
{\Bigg[}\frac{\alpha({\rm K\ {\small I}})}{\alpha({\rm Rb\ {\small I}})}{\Bigg]},
\end{equation}

\noindent where {\it A}{$_g$}(Rb)/{\it A}{$_g$}(K) is the elemental 
abundance ratio, {\it G}(X) is the photoionization rate corrected for 
grain attenuation, and $\alpha$(X) is the rate coefficient for radiative 
recombination. 
We adopted the same sources for the atomic data used by \citet{fed04}
to infer photoionization rates for Rb~{\small I} of 3.42 $\times$ 
10$^{-12}$ s$^{-1}$ and for K~{\small I} of 8.67 $\times$ 10$^{-12}$ s$^{-1}$.
For Rb~{\small I}, the theoretical calculations of \citet{weis} give 
cross sections for 1150-1250 \AA, which were scaled to the 
measurements of \citet{marr} at longer wavelengths.
The cross sections for K~{\small I} are taken from the measurements of 
\citet{hudson1}, \citet{hudson2}, \citet{marr}, and \citet{sandner}.
For wavelengths below 1150 \AA, there is no differential attenuation 
because the ionization potentials are similar (4.18 eV for Rb~{\small I} and
4.34 eV for K~{\small I}).
Based on these cross-sections, a 13\% correction was applied to 
{\it G}(Rb~{\small I}) for $\lambda$ $<$ 1150 \AA.
The rate coefficients for radiative electron recombination are similar, within 
10\% of one another \citep{wane1,peq,wane2}, and so the ratio of the rate 
coefficients is $\sim$ 1.
The ratio of the photoionization rates is 0.394.

For $o$ Per we can solve for {\it N}(Rb~{\small I}), which is 2.51 
$\times$ 10$^{9}$ cm$^{-2}$.
Then, we multiplied this column density by the percentage in each cloud 
[taken from \citet{knauth03}], 14\% and 86\% for the clouds at 3.96 and 
6.98 km s$^{-1}$, respectively. 
Assuming the ratio in each component would be similar to the solar system
value, the relative amounts of the isotopes were separated.
The result was that four starting column densities were created.

The second way to obtain starting column densities was to perform
a one component fit on the data, as in the case of $\zeta$ Oph.
This one component fit provided a coarse estimate of a weighted velocity, 
{\it b}-value, and column density for each isotope.
The isotopes were then split into their
resultant components according to \citet{knauth03}.
The resulting spectra with two components are shown in Figure 2.


\section{Results}

\subsection{The Rubidium Isotope Ratio}

The Rb isotope ratio is expected to be independent of the two principal
effects that hamper the extraction of the Rb/H ratio from the Rb~{\small I}
equivalent widths, namely, the large degree of ionization of Rb to 
Rb$^+$, and the loss of Rb onto grains.
Table 4 shows the final solutions for the lines of sight in our survey.
It lists the resulting velocities, column densities, 
{\it b}-values, $^{85}$Rb/$^{87}$Rb ratios, and reduced $\chi$$^{2}$s, 
calculated from the RbFits program.  The $b$-values range from 0.45 
to 1.46 km s$^{-1}$; while the larger values may indicate the presence of 
unresolved structure, similar values have been reported previously (e.g., 
Knauth et al. 2003).  The reduced $\chi$$^{2}$s are all around 1.0.
The uncertainties for each cloud were calculated using the FWHM, derived
from the {\it b}-value, and S/N of each spectrum.
For such weak lines, the same fractional uncertainty applies to the column
densities also.
These were propagated in the usual way to place uncertainties on the
inferred isotope ratios.
When 2-$\sigma$ was greater than the equivalent widths for the $^{87}$Rb
line, only lower limits to the ratio are possible.
All lines of sight are consistent with the solar system value, 2.59, as 
determined from meteorites, except for HD 147889 and the previously 
measured $\rho$ Oph A (Federman et al. 2004).

We investigated the sensitivity of our results for HD 147889 by attempting
to fit the Rb~{\small I} profile with the solar system ratio. 
Various combinations of keeping $V_{LSR}$ and {\it b}-value fixed were 
attempted.
All gave similar results, with reduced $\chi$$^{2}$s from 1.62 to 1.81 
compared to our preferred synthesis with 1.44.  We performed 
analyses of the Rb isotope ratio toward HD 147889 of a single 
component fit in two additional ways.  
In the first analysis all profile synthesis 
parameters were allowed to vary.  For the second analysis only the 
$^{85}$Rb and $^{87}$Rb column densities were allowed to vary.  For the 
second synthesis, the velocity and the Doppler broadening parameters were 
fixed to those determined from the profile fit to the measured interstellar 
K~{\small I} line.  This led to an improved reduced $\chi^2$ 
over that of the freely varying profile synthesis.  Application of an 
F-test (Lupton 1993) shows the improvement in $\chi^2$ is justified 
to the 68\% confidence level.  
In order to test whether our single component fit of the Rb isotope 
ratio is potentially two components, a subsequent analysis was performed.  
This analysis led to an overall improvement in the reduced $\chi^2$.  
An F-test shows that a second component is justified at the 80\% confidence 
level.  The synthesis based on changing $V_{LSR}$ appears in the lower right 
panel of Figure 1. 
The fit on the blue side is poorer and below the data points, which
is evidenced by the residuals within the left side of the profile 
not appearing random.


\subsection{Comparison with Elemental Potassium Abundance}

Neutral Rb is expected to follow the spatial trend of neutral K in 
interstellar clouds because they have similar ionization potentials and 
chemical properties, and so a determination of the 
interstellar elemental Rb/K ratio from Eqn. 2 compared to the meteoritic 
value may provide new insights into the production of neutron capture elements.
High-resolution surveys have been performed for interstellar 
Li~{\small I} \citep{knauth03}, Na~{\small I} \citep{welty2}, and K~{\small I} 
\citep{welty} absorption.
\citet{knauth03} conclude that the elemental Li/K abundance ratio is 
consistent with the solar system value, and
\citet{welty} also confirm the essentially linear relationship between the
total column densities of Li~{\small I}, Na~{\small I}, and K~{\small I}.
It seems that Li, Na, and K are likely depleted by comparable amounts 
in the relatively cool, dense regions where the neutral species are 
concentrated \citep{welty}.
Thus we apply Eqn. 2 with confidence.

Using this equation Rb/K, as well as $^{85}$Rb/K and $^{87}$Rb/K, were 
determined and the results are presented in Table 5.
The results for $\rho$ Oph A \citep{fed04} and the solar system values 
\citep{lodders} are shown for comparison.
The K~{\small I} column densities were obtained from the weak line 
at 4044 \AA, except for AE Aur, where we adopted the unpublished 
$\lambda$4047 results of P. Boiss\'{e}.
The K~{\small I} value for HD 147889 comes from \citet{welty}.

The seven new lines of sight given in Table 5 all show a deficit in the 
ratios relative to potassium, with the
possible exception of $^{87}$Rb/K toward $\rho$ Oph A.
Relative to the meteoritic ratios, we find Rb/K=34\%, $^{85}$Rb/K=32\%, and 
$^{87}$Rb/K=36\%.  These ratios are inferred from Eqn. (2), which does 
not allow for different depletions onto grains.  As mentioned below, 
the condensation temperature for Rb is lower than that for K; if only 
depletion mattered, one would expect a higher Rb/K ratio, contrary to the 
observational results.  
Note that $^{85}$Rb/K toward $o$ Per seems to be larger than the other 
ratios, and that the anomalously large ratios of $^{87}$Rb/K (HD 147889 
and $\rho$ Oph A) are not included in the above averages.

\section{Discussion}

\subsection{Comparison with Previous Results}

The first limits on interstellar {\it N}(Rb~{\small I}) come from \citet{fed85}, who 
searched for absorption toward the stars $o$ Per, $\zeta$ 
Per, and $\zeta$ Oph, but did not detect any to a 3-$\sigma$ limit of 
$\leq$ 1.5 m\AA.
The resulting limits on {\it N}(Rb~{\small I}) and Rb/K are found to be 
typically $\leq$ 5 $\times$ 10$^{9}$ cm$^{-2}$ and $\leq$ 1.5 $\times$ 
10 $^{-3}$ \citep{fed04}.
In all cases, our detections yield values 30 to 50\% of the upper limits.

\citet{gredel} found column densities toward Cyg OB2 No. 5 and No. 12 of 
{\it N}(Rb~{\small I}) $=$ (7 $\pm$ 2) $\times$ 10$^{9}$ cm$^{-2}$ and 
{\it N}(Rb~{\small I}) $=$ (13 $\pm$ 2) $\times$ 10$^{9}$ cm$^{-2}$, 
respectively.
Combining these results with the high-resolution K~{\small I} $\lambda$7699
spectra of \citet{mccall}, \citet{fed04} derived values for 
{\it N}(K~{\small I}) of the three main molecular components to be
9.4(7.7) $\times$ 10$^{11}$, 8.1(12) $\times$ 10$^{11}$, and
6.9(16) $\times$ 10$^{11}$ cm$^{-2}$ for the gas toward Cyg OB2 No. 5(12).
Thus, Rb/K is 14 $\times$ 10$^{-4}$ for No. 5 and 
12 $\times$ 10$^{-4}$ for No. 12.
Rb/K for $\rho$ Oph A is (13 $\pm$ 3) $\times$ 10$^{-4}$ \citep{fed04}.
These three lines of sight appear to have Rb/K values about a factor of two
higher than the 
average of our eleven clouds, (6.1 $\pm$ 1.2) $\times$ 10$^{-4}$. 
However, significant uncertainty remains in extracting elemental abundances
toward the stars in Cyg OB2 because a blended, optically thick line was used
to obtain the K~{\small I} column density.
Our analyses, and those of \citet{snow}, for HD 147889 indicate that the 
weaker line at 4044 \AA\ yields a higher K~{\small I} column density, leading
to a smaller Rb/K ratio.
Furthermore, the difference for the gas toward $\rho$ Oph A is within 
$\sim$2-$\sigma$ of the results for our other lines of sight.
Thus, all interstellar results appear smaller than the solar system value
of (18~$\pm$~4) $\times$ 10$^{-4}$.

From the Rb~{\small I} detection toward $\rho$ Oph A,  
\citet{fed04} suggested that there was a deficit of r-processed material 
because $^{87}$Rb/K seemed to have the meteoritic value, while $^{85}$Rb/K
was considerably less than the solar system value.
This is also true toward HD 147889. 
These two lines of sight are near one another; the anomalously low 
$^{85}$Rb/$^{87}$Rb ratios seem to arise from sampling the Rho Ophiuchus 
Molecular Cloud.
For the other lines of sight, our sample indicates that both $^{85}$Rb 
and $^{87}$Rb have a deficit relative to K in the ISM of $\sim$66\%. 
However, the gas in the Rho Ophiuchus Molecular Cloud appears to be 
enriched in $^{87}$Rb.

Kawanomoto et al. (2009) recently published results 
for the direction toward HD~169454.  Their isotope ratio 
($^{85}$Rb/$^{87}$Rb $>$ 2.4) is consistent with the meteoritic value, which 
again highlights the unusual values that we found for the Rho Ophiuchus 
Molecular Cloud.  Their Rb/K ratio also reveals a deficit 
compared to the meteoritic ratio.  It is worth noting that Kawanomoto 
et al. make comparisons with a solar system ratio at its formation by 
accounting for the decay of $^{87}$Rb over 4.55 Gyr.  Their estimate of 
the original ratio is 2.43.

\subsection{Interstellar Results on Neutron-Capture Elements}

The elemental abundances of 
Ge, Kr, Cd, and Sn seen in our sample of diffuse clouds, 
for which sufficient data exist, may provide an answer, 
because they are also synthesized through neutron capture. 
For each of these elements, the dominant ion is observed.
In the solar system about 10\% of Ge, 20\% of Kr, 50\% of Cd,
and 70\% of Sn are produced via the main s-process
\citep{beer,raiteri,arland,heil}; massive stars through a combination of weak
s-process and r-process are responsible for the remaining fractions.
When corrected for the {\it f}-value adopted by \citet{cart06}, 
\citet{cardelli91} obtain {\it N}(Ge)/{\it N}(H) of 
(4.9~$\pm$~0.5) $\times$ 10$^{-10}$ for $\zeta$ Oph,
while \citet{lodders} gives (4.2~$\pm$~0.5) $\times$ 10$^{-9}$ for the solar 
system.
\citet{cart} find {\it N}(Kr)/{\it N}(H) to be (9.33~$\pm$~1.14) $\times$ 
10$^{-10}$ for $\zeta$ Per and (7.94~$\pm$~0.97) $\times$ 10$^{-10}$ for 
$\zeta$ Oph, whereas the theoretical solar system abundance from 
\citet{lodders} is (19.05~$\pm$~3.81) $\times$ 10$^{-10}$.
Sn was measured by \citet{sofia}, who found {\it N}(Sn)/{\it N}(H) of 
(7.41~$\pm$~3.18) $\times$ 10$^{-11}$ for $\zeta$ Per, (8.13~$\pm$~0.98) 
$\times$ 10$^{-11}$ for $\chi$ Oph, (9.12~$\pm$~2.65) $\times$ 10$^{-11}$ 
for $\zeta$ Oph, and (5.75~$\pm$~0.95) $\times$ 10$^{-11}$ for $\rho$ Oph A.
The solar system abundance is (12.88~$\pm$~1.24) $\times$ 10$^{-11}$ 
\citep{lodders}.
Finally, there is one line of sight among our sample that has measured Cd, 
$\zeta$ Oph, where {\it N}(Cd)/{\it N}(H) is (5.50~$\pm$~1.60) $\times$ 
10$^{-11}$ \citep{sofia}.
The solar system value for this element is (5.50~$\pm$~0.38) $\times$
10$^{-11}$ \citep{lodders}.
Taking into account the uncertainties, all these values are likely to be 
lower than the solar system values.

For these relatively dense lines of sight, depletion onto grains, not
nucleosynthetic processes, could be the cause for the deficits.
It is better to consider the gas phase abundances for low density, warm gas.
The condensation temperatures for Ge, Kr, Cd, Sn, and Rb are 734, 52, 652, 704, 
and 800 K, respectively \citep{lodders}.
The survey by \citet{cart06} indicates an interstellar Ge abundance that is 
25$\%$ the meteoritic value.
With such a low condensation temperature, Kr should not be depleted toward
any line of sight.
However, the average abundance of Kr found by \citet{cart} is $\sim$50$\%$ 
of the solar system value.
The results on Cd and Sn \citep{sofia} reveal a different trend.
For low density lines of sight, neither element shows any depletion onto 
grains.
There is even a hint that the Sn abundance for these directions is greater
than the solar system value.
The deficit of Rb relative to K cannot be explained by depletion onto grains 
resulting from a
higher condensation temperature, because Li, Na, and K have even higher 
values (1135, 953, and 1001 K) according to \citet{lodders}, and the Li/K 
and Na/K elemental ratios are similar to meteoritic values \citep{welty,knauth03}.

The depletion patterns of Ge, Kr, and Rb on the one hand, and Cd and Sn on 
the other, cannot be attributed to imprecise oscillator strengths. 
According to \citet{morton03}, the $f$-values used to study interstellar 
Kr~{\small I}, Rb~{\small I}, and Cd~{\small II} are well known from 
laboratory measurements. 
The same applies to Sn~{\small II} as a result of the experimental results 
of \citet{schectman}. 
\citet{morton03} recommends the theoretical value of \citet{biemont} for 
the Ge~{\small II} line at 1237 \AA. 
While no more recent theoretical or experimental determinations are 
available at the present time, the dichotomy noted above would still be 
present if the results on interstellar Ge were not included.

A summary of the depletion patterns for the five elements appears in the top 
panel of Fig. 3. 
Observational results for the amount of depletion onto grains both for 
low and high density clouds is shown.
Each survey used to construct this figure [\citet{cart} for Kr; 
\citet{cart06} for Ge; our results for Rb; \citet{sofia} for Cd and Sn]
indicate typical observational scatter of about 0.1 dex in the quoted depletion.
Although $T_{cond}$ is similar for Ge, Rb, Cd, and Sn, the depletion 
patterns for Cd and Sn are strikingly different than those for Ge and Rb, 
even when the intrinsic scatter in plots of depletion versus 
$T_{cond}$ are considered.  
Furthermore, there is less depletion for Cd and Sn than for Kr, a noble
gas.

A closer look at the production routes for these n-capture nuclides offers
a promising solution.
This is highlighted in the bottom panel of Fig. 3, where the white bar indicates
production by low-mass stars and the shaded bar indicates massive stars.
The solar system abundances for the elements Cd and Sn are believed to arise
predominantly by the main s-process involving low-mass AGB stars
\citep{beer,raiteri,arland}, and they have reduced levels of depletion
in the ISM.
On the other hand, elements synthesized by the weak s-process and the 
r-process that occur in massive stars 
\citep{beer,raiteri,arland,the,heil} 
appear to show interstellar abundances 
significantly lower than seen in the solar system.
This possible reduction in the contribution of massive stars to interstellar
abundances for n-capture elements contrasts with recent results for lighter
elements.
For instance, \citet{przy} provide abundances for several elements in 
unevolved, early-type B stars that are indistinguishable from those found
for the Orion Nebula \citep{esteban} and the sun \citep{asplund}.
The enhancement of $^{87}$Rb in the Rho Ophiuchus Molecular Cloud remains 
puzzling.
Our inferences may not be appropriate if Galactic chemical evolution 
altered the mix of s- and r-processed material from 4.5 Gyr ago when the 
solar system formed to the ISM of today.
While we are not aware of predictions of the evolution of Rb and its 
isotopes, the weak s-process and r-process from Type II SN might be tied 
to the evolution of O or Mg or Si, while the main s-process to products
of AGB stars such as Y or Ba.

Further efforts in a number of areas will likely improve our understanding.
Data on interstellar Ga, As, and Pb -- other n-capture nuclides -- are 
available, but as of now it is more difficult to disentangle the effects of 
nucleosynthesis from depletion onto grains for them.
Once this is accomplished, the unexpected results on Rb and its isotopes
will likely be clarified.
A more complete set of theoretical results, covering all isotopes of 
interest and based on modern stellar codes \citep[e.g.,][]{the,heil}, are 
needed as well.
We also see a need for theoretical studies into the evolution of these 
elements and their isotopes.

\section{Summary and Conclusions}

The interstellar $^{85}$Rb/$^{87}$Rb ratio was measured along seven 
lines of sight, $o$ Per, $\zeta$ Per, AE Aur, HD 147889, $\chi$ Oph, 
$\zeta$ Oph, and 20 Aql.
Using K~{\small I} absorption as a guide, the spectra were synthesized 
to find the velocity, column density, and 
{\it b}-value of each cloud. 
Most clouds had ratios that agreed, within mutual uncertainties, 
with the solar system value of 2.59.
The gas toward HD 147889 had a lower $^{85}$Rb/$^{87}$Rb ratio 
(1.03~$\pm$~0.21) and a higher elemental ratio (8.6~$\pm$~1.3),
suggesting that $^{87}$Rb is enhanced in this direction, consistent
with the previous results for $\rho$ Oph A \citep{fed04}.
The anomalous ratios appear to arise within the Rho Ophiuchus Molecular Cloud. 
Furthermore, a comparison of interstellar Rb to K shows that the gas 
in the solar neighborhood appears to be significantly underabundant in 
rubidium, even after considering uncertainties in ionization balance and 
level of depletion.
An examination of other species, such as Ge, Kr, Cd, and Sn, seems to suggest 
nuclides synthesized by the n-capture processes occuring in massive stars 
(weak s and r) have lower interstellar abundances than expected.
Larger samples for several n-capture elements, from both observation and theory, 
will help clarify the cause for this.

\acknowledgments

We thank Dr. Y. Sheffer for his help with his program ISMOD.
We acknowledge the use of the DIB database of Professor D. York at the
University of Chicago, from which promising directions were noted.
We also thank Prof. P. Boiss\'{e} for the AE Aur K~{\small I} $\lambda$4047 
spectrum.
DLL thanks the Robert A. Welch Foundation of Houston for their support 
through grant number F-634.
We made use of the SIMBAD database, operated at Centre de Don\'{e}es
Astronomiques de Strasbourg, Strasbourg, France.


\clearpage

\begin{figure}
\epsscale{1.0}
\plotone{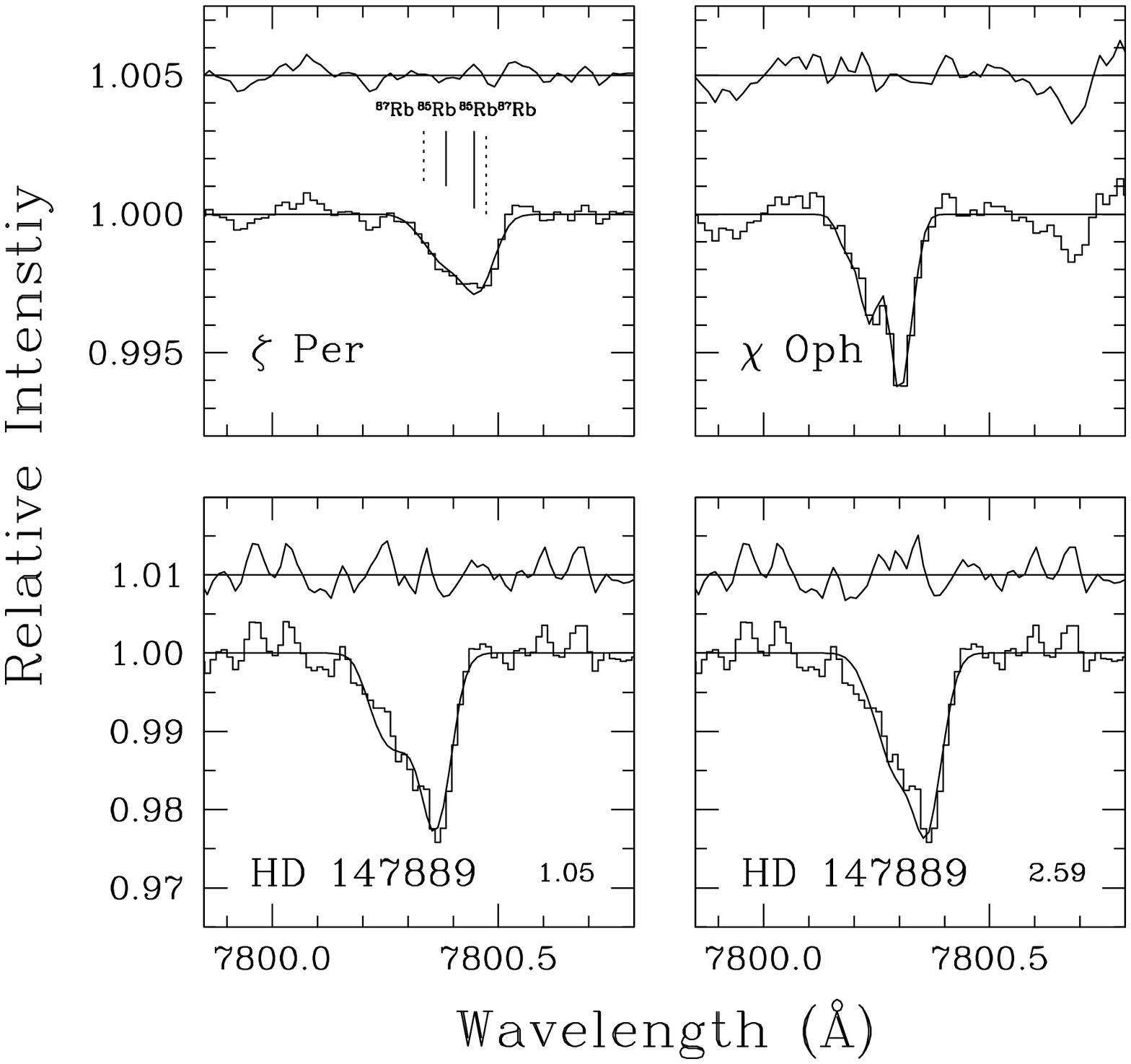}
\caption[Synthesis of Rb~{\small I} toward $\zeta$ Per, $\chi$ Oph, and HD 
147889.]{Synthesis of Rb~{\small I} toward $\zeta$ Per, $\chi$ Oph, and HD 
147889 based on one-component fits. 
The histogram indicates the data, while the solid line is the fit. 
The residuals (data minus fit) are offset to 1.005, except for HD 147889, 
where they are offset to 1.01. The vertical solid lines above the
spectrum of $\zeta$ Per indicate $^{85}$Rb 
while the dashed lines indicate $^{87}$Rb. The lengths show the relative 
line strengths. The feature in $\chi$ Oph at 7800.68 \AA\ 
is a glitch that did not compromise our spectrum. Note the expanded scale 
for HD 147889. The bottom left panel is fit with a $^{85}$Rb/$^{87}$Rb ratio 
of 1.05, while the bottom right panel is fit with 
the solar system ratio; the isotope ratio is given in the lower right corner.}
\end{figure}

\clearpage

\begin{figure}
\epsscale{1.0}
\plotone{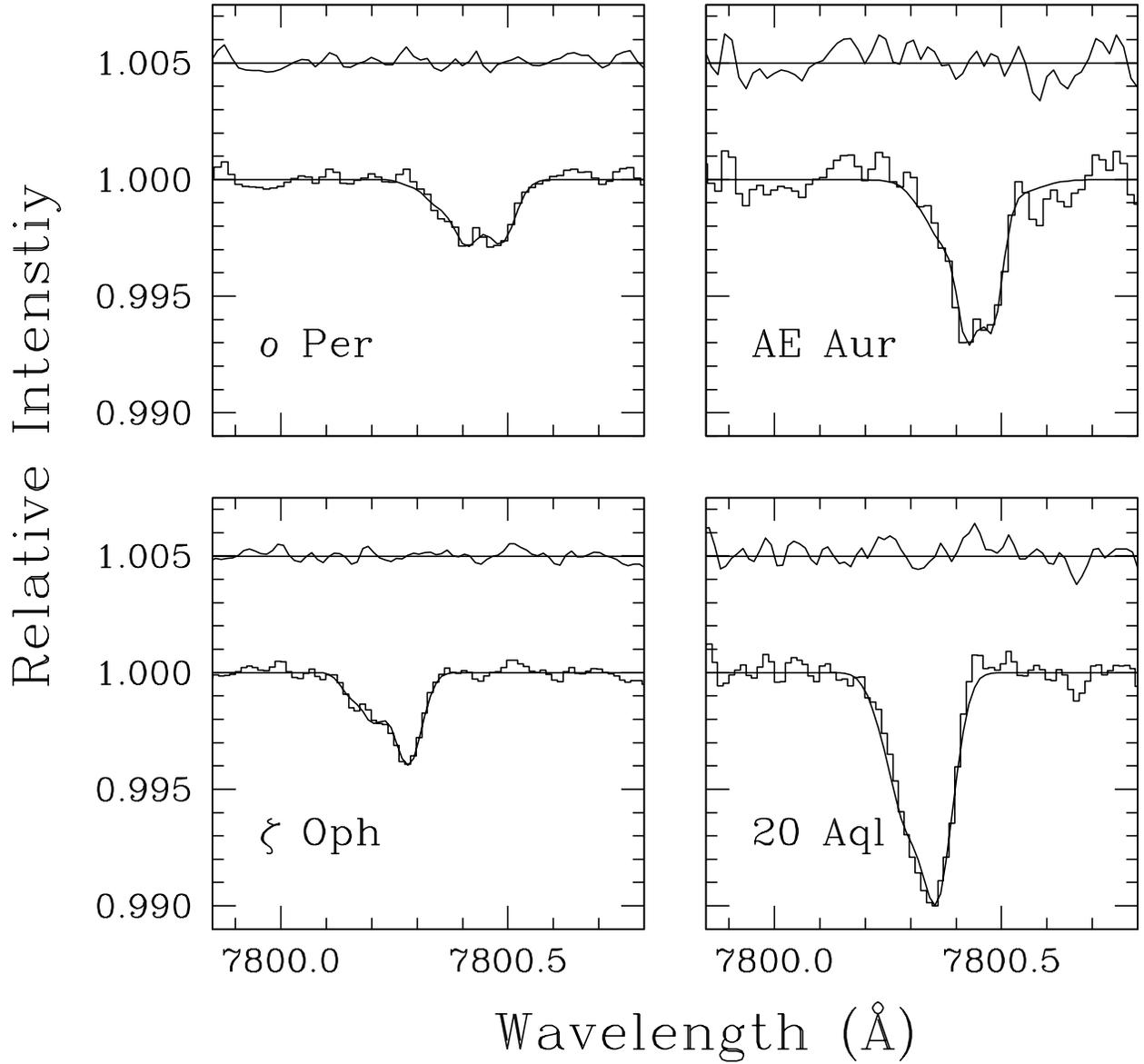}
\caption[Synthesis of Rb~{\small I} toward $o$ Per, AE Aur, $\zeta$ Oph and 
20 Aql.]{Synthesis of Rb~{\small I} toward $o$ Per, AE Aur, $\zeta$ Oph and 
20 Aql based on two-component fits. Same as Figure 1.}
\end{figure}

\clearpage

\begin{figure}
\epsscale{1.0}
\plotone{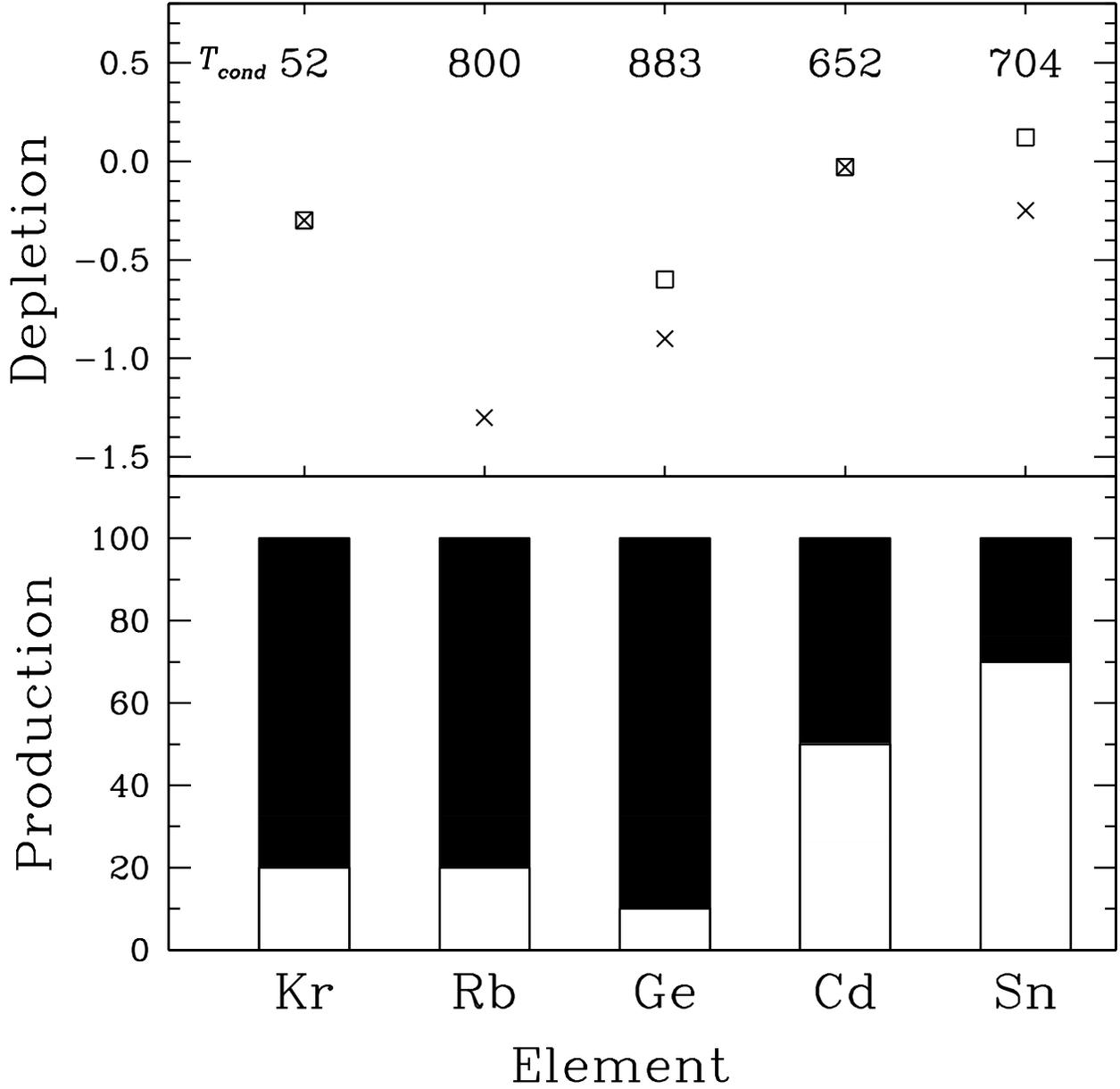}
\caption[Production Sites]{The upper panel depicts depletion for warm 
gas (squares) and cold gas (x's) for Ge, Kr, Rb, Cd, and Sn along with their
condensation temperatures. The uncertainties for the measured abundances 
are smaller than the symbols.  The two groupings in production 
mainly from massive stars and mainly from low-mass stars are 
ordered by condensation temperature. The bottom panel
shows production from massive stars (shaded bars) and low mass 
stars (white bars).}
\end{figure}

\clearpage

\begin{deluxetable}{rrcccccc}
\tablewidth{0pt}
\rotate
\tablecaption{Observational Data\label{Obsdata1-Table}}
\tablehead{
\colhead{HD} & \colhead{Name} & \colhead{Type} & \colhead{${\it V}$} & 
\colhead{${\it d}$} & \colhead{Date Observed} & \colhead{Exposure Time} & 
\colhead{Unreddened}\\ 
\colhead{} & \colhead{} & \colhead{} & \colhead{(mag)} & \colhead{(pc)} & 
\colhead{} & \colhead{(min)} & \colhead{Star}}
\startdata
23180  & $o$ Per & B1 III & 4.98 & 450 & Dec. 2005, Jan. 2007 & 1470 &
$\alpha$ Leo, $\gamma$ Cas \\
24398  & $\zeta$ Per & B1 Iab & 2.88 & 300 & Dec. 2005, Jan. 2007 & 946 &
$\alpha$ Leo, $\gamma$ Cas \\
34078  & AE Aur & O9.5 Ve & 6.00 & 450 & Dec. 2005, Jan. 2007 & 1500 &
$\alpha$ Leo, $\gamma$ Cas \\
147889 & \ldots & B2 III/IV & 7.95 & 135 & Jun. 2006 & 560 & 
$\alpha$ Vir \\
148184 & $\chi$ Oph & B2 V & 4.42 & 150 & May 2005 & 810 &
$\alpha$ Vir, $\alpha$ Lyr \\
149757 & $\zeta$ Oph & O9 V & 2.58 & 140 & May 2005, Jun. 2006 & 801 &
$\alpha$ Vir, $\alpha$ Lyr \\
179406 & 20 Aql & B3 V & 5.36 & 375 & May 2005, Jun. 2006 & 1055 &
$\alpha$ Vir, $\alpha$ Lyr \\
\enddata
\end{deluxetable}

\clearpage

\begin{deluxetable}{ccc}
\tablecaption{Rb~{\small I} Spectra\label{RbIspectra2-Table}}
\tablewidth{0pt}
\tablehead{
\colhead{Name} & \colhead{S/N} & \colhead{{\it W}$_\lambda$/$\sigma$}}
\startdata
$o$ Per & 3300 & 8.9 \\
$\zeta$ Per & 4500 & 10.8 \\
AE Aur & 1500 & 10.5 \\
HD147889 & 700 & 13.2 \\
$\chi$ Oph & 2100 & 11.6 \\
$\zeta$ Oph & 4100 & 16.9 \\
20 Aql & 2100 & 18.4 \\
\enddata
\end{deluxetable}

\clearpage

\begin{deluxetable}{ccc}
\tablecaption{Rb~{\small I} Hyperfine Structure Component Wavelengths and 
{\it f}-values\label{RbIfvalues3-Table}}
\tablewidth{0pt}
\tablehead{
\colhead{Species} & \colhead{$\lambda$(\AA)} & \colhead{{\it f}-value}}
\startdata
$^{85}$Rb & 7800.232 & 0.290 \\
          & 7800.294 & 0.406 \\
$^{87}$Rb & 7800.183 & 0.261 \\
          & 7800.321 & 0.435 \\
\enddata
\end{deluxetable}

\clearpage

\begin{deluxetable}{lcccccc}
\tablecaption{Isotope Ratios\label{isoratios4-Table}}
\tablewidth{0pt}
\tabletypesize{\footnotesize}
\tablehead{
\colhead{Star}     & \colhead{Isotope}       & \colhead{{\it V}$_{LSR}$} & 
\colhead{{\it N}} & \colhead{{\it b}-value} & 
\colhead{$^{85}$Rb/$^{87}$Rb}                & 
\colhead{Reduced} \\
\colhead{}         & \colhead{}              & \colhead{(km s$^{-1}$)}   & 
\colhead{(cm$^{-2}$)} & \colhead{(km s$^{-1}$)}	& \colhead{Ratio} & 
\colhead{$\chi$$^{2}$}}
\startdata
$o$ Per	&	87	&	4.05	&	$<$1.00 $\times$ 10$^8$	&
	0.84	&		&		\\
	&	85	&	4.05	&	(2.72 $\pm$ 0.35) $\times$ 10$^8$	&	
0.84	&	$>$2.72	&		\\
	&	87	&	7.07	&	(2.70 $\pm$ 0.30) $\times$ 10$^8$	&	
1.06	&			&		\\
	&	85	&	7.07	&	(6.40 $\pm$ 0.71) $\times$ 10$^8$	&	
1.06	&	2.37 $\pm$ 0.79	&	0.72	\\
	&		&		&	&	&	&		\\
$\zeta$ Per	&	87	&	5.88	&	(3.29 $\pm$ 0.29) $\times$ 10$^8$	&	
1.46		&		&		\\ 
	&	85	&	5.88	&	(7.49 $\pm$ 0.67) $\times$ 10$^8$	&	
1.46	&	2.28 $\pm$ 0.59	&	1.75	\\
	&		&		&	&	&	&		\\
AE Aur	&	87	&	5.34	&	$<$3.00 $\times$ 10$^8$	&
1.03		&		&		\\
	&	85	&	5.34	&	(8.23 $\pm$ 1.07) $\times$ 10$^8$	&	
1.03	&	$>$2.74		&		\\
	&	87	&	7.30	&	(4.50 $\pm$ 0.59) $\times$ 10$^8$	&	
1.00	&			&		\\
	&	85	&	7.30	&	(1.13 $\pm$ 0.15) $\times$ 10$^9$	&	
1.00	&	2.52 $\pm$ 1.03	&	1.06	\\
	&		&		&	&	&	&		\\
HD 147889	&	87	&	2.04	&	(4.21 $\pm$ 0.35) $\times$ 10$^9$	&	
1.33		&		&		\\
	&	85	&	2.04	&	(4.33 $\pm$ 0.36) $\times$ 10$^9$	&	
1.33	&	1.03 $\pm$ 0.21	&	1.44	\\
	&		&		&	&	&	&		\\
$\chi$ Oph	&	87	&	0.13	&	(4.85	$\pm$ 0.38) $\times$ 10$^8$ &	
0.46		&		&		\\
	&	85	&	0.13	&	(1.27 $\pm$ 0.11) $\times$ 10$^9$	&	
0.46	&	2.62 $\pm$ 0.63	&	2.06	\\
	&		&		&	&	&	&		\\
$\zeta$ Oph	&	87	&	-1.21	&	(2.99 $\pm$ 0.25) $\times$ 10$^8$	&	
0.45		&		&		\\
	&	85	&	-1.21	&	(4.72 $\pm$ 0.39) $\times$ 10$^8$	&	
0.45	& 1.58 $\pm$ 0.32	&		\\
	&	87	&	-0.17	&	(1.54	$\pm$ 0.21) $\times$ 10$^8$   &	
0.59		&		&		\\
	&	85	&	-0.17	&	(3.19 $\pm$ 0.43) $\times$ 10$^8$	&	
0.59	& 2.07 $\pm$ 0.78	&	0.94	\\
	&		&		&	&	&	&		\\
20 Aql	&	87	&	2.06	&	(7.26 $\pm$ 0.53) $\times$ 10$^8$	&	
1.28		&		&		\\
	&	85	&	2.06	&	(2.05 $\pm$ 0.15) $\times$ 10$^9$	&	
1.28	&	2.82 $\pm$ 0.68	&		\\
	&	87	&	3.04	&	$<$2.42 $\times$ 10$^8$	&	
1.19	&			&		\\
	&	85	&	3.04	&	(6.06 $\pm$ 0.31) $\times$ 10$^8$	&	
1.19	&	$>$2.51		&	0.90	\\
\enddata
\end{deluxetable}

\clearpage

\begin{deluxetable}{lrcccc}
\tablecaption{Abundance Ratios from the Neutral Species\label{abundratios-Table}}
\tablewidth{0pt}
\tablehead{									
\colhead{Star} & \colhead{{\it V}$_{LSR}$} & \colhead{$^{85}$Rb/$^{87}$Rb} & 
\colhead{Rb/K\tablenotemark{a}}  & 
\colhead{$^{85}$Rb/K\tablenotemark{a}} & 
\colhead{$^{87}$Rb/K\tablenotemark{a}}\\
\colhead{} & \colhead{(km s$^{-1}$)} & \colhead{} & \colhead{} & \colhead{} 
& \colhead{}}
\startdata
$o$ Per	&	4.05	&	$>$2.72	&	10.0 $\pm$ 1.0	&	
7.3 $\pm$ 1.3	&	$<$2.7	\\
	&	7.07	&	2.37 $\pm$ 0.79	&	4.0 $\pm$ 1.1	&	
2.8 $\pm$ 0.7 & 1.2 $\pm$ 0.4		\\
\\
$\zeta$ Per	&	5.88	&	2.28 $\pm$ 0.59	&	6.1 $\pm$ 0.6	&
4.2 $\pm$ 0.3	&	1.9 $\pm$ 0.3	\\
\\
AE Aur	&	5.34	&	$>$2.74	&	3.0 $\pm$ 0.6\tablenotemark{b} &	
2.2 $\pm$ 0.8\tablenotemark{b} 	&	$<$0.8\tablenotemark{b} 		\\
	&	7.30	&	2.52 $\pm$ 1.03	&	6.1 $\pm$ 1.3\tablenotemark{b} &
4.3 $\pm$ 0.9\tablenotemark{b}  & 1.7 $\pm$ 0.4\tablenotemark{b}   \\
\\
HD 147889	&	2.04	&	1.03 $\pm$ 0.21	&	
8.6 $\pm$ 1.3\tablenotemark{c} 	& 4.4 $\pm$ 0.7\tablenotemark{c} 	& 
4.3 $\pm$ 0.6\tablenotemark{c} 		\\
\\
$\chi$ Oph	&	0.13	&	2.62 $\pm$ 0.63	&	5.4 $\pm$ 0.7	&
3.9 $\pm$ 0.4	&	1.5 $\pm$ 0.3	\\
\\
$\zeta$ Oph	&	-1.21	&	1.58 $\pm$ 0.32	&	5.7 $\pm$ 0.6
	&	3.5 $\pm$ 0.3	&	2.2 $\pm$ 0.3	\\
	&	-0.17	&	2.07 $\pm$ 0.78	&	6.4 $\pm$ 1.1	&
	4.3 $\pm$ 0.7	&	2.1 $\pm$ 0.5	\\
\\
20 Aql	&	2.06	&	2.82 $\pm$ 0.68	&	7.0 $\pm$ 0.7	&
5.2 $\pm$ 0.5	&	1.8 $\pm$ 0.3	\\
& 3.04	&	$>$2.51	&	5.2 $\pm$ 0.7	&	3.7 $\pm$ 0.8	&
	$<$1.5	\\
\\
$\rho$ Oph A & \nodata & 1.21 $\pm$ 0.30\tablenotemark{d} & 
13.0 $\pm$ 3.0\tablenotemark{d}& 7.3 $\pm$ 1.2\tablenotemark{d} &	
6.1 $\pm$ 1.3\tablenotemark{d} \\
\\
Solar System	& \nodata &	2.59\tablenotemark{e}	&	
17.8 $\pm$ 3.7\tablenotemark{e} & 12.9\tablenotemark{e} & 4.9\tablenotemark{e} \\
\enddata
\tablenotetext{a}{The values for Rb/K, $^{85}$Rb/K, and $^{87}$Rb/K are 
in units of 10$^{-4}$.}
\tablenotetext{b}{From data provided by P. Boiss\'{e}.}
\tablenotetext{c}{K~{\small I} values are from \citet{welty}.}
\tablenotetext{d}{\citet{fed04}.} 
\tablenotetext{e}{\citet{lodders}.}						
\end{deluxetable}

\end{document}